\documentstyle[12pt,epsf]{article}

\newcommand{\beq}{\begin{equation}}
\newcommand{\eeq}{\end{equation}}

\newcommand{\subhead}[1]{\goodbreak \vskip 3.25ex plus
   1ex minus 0.2ex \noindent {\large \bf
   #1}\par\nobreak \vskip 1.5ex plus 0.2ex}

     \def\lta{\mathrel{\vcenter{\vbox{\offinterlineskip \hbox{$<$}
          \vskip 0.2 pt \hbox{$\sim$}}}}}
     \def\tot{{\rm tot}}

\begin{document}

\begin{titlepage}

\setcounter{footnote}{1}
\setcounter{page}{1}

\vbox{}
\flushright{MIT-CTP-2947, astro-ph/0002188}
\vskip 1.5 truein
\begin{center}
{\Large \bf Inflationary Models\\}
\medskip
{\Large \bf and\\}
\medskip
{\Large \bf Connections to Particle
Physics\footnote{Talk given at the {\it Pritzker Symposium on the
Status of Inflationary Cosmology}, Chicago, Illinois, January 29
-- 31, 1999.  To appear in the proceedings.}\\}
\bigskip \bigskip
{\bf Alan H. Guth\footnote{This work is supported in part by
funds provided by the U.S. Department of Energy (D.O.E.) under
cooperative research agreement \#DF-FC02-94ER40818, and in part
by funds provided by NM Rothschild \& Sons Ltd and by the
EPSRC.}\\}
\medskip
{\small \it Center for Theoretical Physics\\}
{\small \it Laboratory for Nuclear Science and
Department of Physics\\}
{\small \it Massachusetts Institute of Technology,
Cambridge, Massachusetts\ \ 02139\ \ \ U.S.A.\footnote{Present
address.}\\ and \\
Isaac Newton Institute for Mathematical Sciences, \\
Clarkson Road, Cambridge CB3 0EH, UK\\}
\bigskip
{\tt guth@ctp.mit.edu\\}
\bigskip \bigskip
\end{center}

\begin{abstract}
The basic workings of inflationary models are summarized, along
with the arguments that strongly suggest that our universe is the
product of inflation.  The mechanisms that lead to eternal
inflation in both new and chaotic models are described.  Although
the infinity of pocket universes produced by eternal inflation
are unobservable, it is argued that eternal inflation has real
consequences in terms of the way that predictions are extracted
from theoretical models.  The ambiguities in defining
probabilities in eternally inflating spacetimes are reviewed,
with emphasis on the youngness paradox that results from a
synchronous gauge regularization technique.  To clarify (but not
resolve) this ambiguity, a toy model of an eternally inflating
universe is introduced.  Vilenkin's proposal for avoiding these
problems is also discussed, as is the question of whether it is
meaningful to discuss probabilities for unrepeatable
measurements.
\end{abstract}

\end{titlepage}

\setcounter{footnote}{0}

\section{Introduction}
\setcounter{equation}{0}

There are many fascinating issues associated with eternal
inflation, which will be the main subject of this talk. You have
certainly heard other people talk about eternal inflation, but I
feel that the topic is important enough so that you should hear
about it in some accent other than Russian.  I will begin by
summarizing the basics of inflation, including a discussion of
how inflation works, and why many of us believe that our universe
almost certainly evolved through some form of inflation.  This
material is certainly not new, but I think it is an appropriate
introduction to any volume that focuses on inflationary
cosmology.  Then I will move on to discuss eternal inflation,
first explaining how it works.  I will then argue the eternal
inflation has important implications, and raises important
questions, which should not be dismissed as being merely
metaphysical.

\section{How Does Inflation Work?}
\setcounter{equation}{0}

In this section I will review the basics of how inflation works,
focusing on the earliest working forms of inflation---{\it new
inflation} \cite{Linde1, Albrecht-Steinhardt1} and {\it chaotic
inflation} \cite{chaotic}.  While more complicated possibilities
(e.g. hybrid inflation \cite{hyb1,hyb2,hyb3,hyb4,hyb5} and
supernatural inflation \cite{RSG}) appear very plausible, the
basic scenarios of new and chaotic inflation will be sufficient
to illustrate the physical effects that I want to discuss in this
article. 

The key property of the laws of physics that makes inflation
possible is the existence of states of matter that have a high
energy density which cannot be rapidly lowered.  In the original
version of the inflationary theory \cite{Guth1}, the proposed
state was a scalar field in a local minimum of its potential
energy function.\footnote{A similar proposal was advanced by
Starobinsky \cite{Starobinsky}, in which the high energy density
state was achieved by curved space corrections to the
energy-momentum tensor of a scalar field.}  Such a state is
called a {\it false vacuum}, since the state temporarily acts
as if it were the state of lowest possible energy density. 
Classically this state would be completely stable, because there
would be no energy available to allow the scalar field to cross
the potential energy barrier that separates it from states of
lower energy.  Quantum mechanically, however, the state would
decay by tunneling \cite{Coleman}.  Initially it was hoped that
this tunneling process could successfully end inflation, but it
was soon found that the randomness of false vacuum decay would
produce catastrophically large inhomogeneities \cite{Guth1, HMS,
GuthWeinberg}.  

\begin{figure}[ht]
\epsfxsize=201pt 
\centerline{\epsfbox{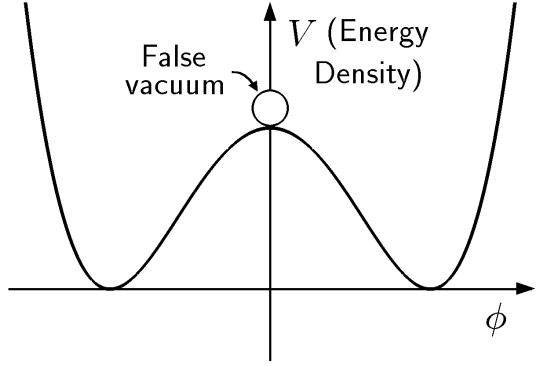}}
\caption{Generic form of the potential for the new inflationary
scenario.} 
\label{newinf}
\end{figure}

This ``graceful exit'' problem was solved by the invention of the
new inflationary universe model \cite{Linde1,
Albrecht-Steinhardt1}, which achieved all the successes that had
been hoped for in the context of the original version.  In this
theory inflation is driven by a scalar field perched on a plateau
of the potential energy diagram, as shown in Fig.~\ref{newinf}. 
Such a scalar field is generically called the {\it inflaton}.  If
the plateau is flat enough, such a state can be stable enough for
successful inflation. Soon afterwards Linde showed that the
inflaton potential need not have either a local minimum or a
gentle plateau; in chaotic inflation \cite{chaotic}, the
inflaton potential can be as simple as
\beq
   V(\phi)={1 \over 2} m^2 \phi^2, 
   \label{eq:2.1}
\eeq
provided that $\phi$ begins at such a large value that it takes a
long time for it to relax.  For simplicity of language, I will
stretch the meaning of the phrase ``false vacuum'' to include all
of these cases; that is, I will use the phrase to denote any
state with a high energy density that cannot be rapidly
decreased.  While inflation was originally developed in the
context of grand unified theories, the only real requirement on
the particle physics is the existence of a false vacuum state.

\subhead{The New Inflationary Scenario:}

Suppose that the energy density of a state is approximately equal
to a constant value $\rho_f$.  Then, if a region filled with this
state of matter expanded by an amount $dV$, its energy would have
to increase by
\beq
  d U = \rho_f \, d V \ . 
  \label{eq:2.2}
\eeq
Something would have to supply that energy.  Work would have to
be done to cause the region to expand, which implies that the
region has a negative pressure, which pulls back against whatever
is causing the region to expand.  The work done by this negative
pressure $p_f$ is given by the elementary formula
\beq
  dW = - p_f \, d V \ .
  \label{eq:2.3}
\eeq
Equating the work with the change in energy, one finds
\beq
  p_f = - \rho_f \ . 
  \label{eq:2.4}
\eeq
It is this negative pressure which is the driving force behind
inflation.  When one puts this negative pressure into Einstein's
equations to find out its gravitational effect, one finds that it
leads to a repulsion, causing such a region to undergo
exponential expansion.  If the region can be approximated as
isotropic and homogeneous, this result can be seen from the
standard Friedmann-Robertson-Walker (FRW) equations:
\beq
  {d^2 a \over d t^2} = - {4 \pi \over 3} G ( \rho + 3 p ) a \ 
   = { 8 \pi \over 3 } G \rho_f a  \ .
  \label{eq:2.5}
\eeq
where $a(t)$ is the scale factor, $G$ is Newton's constant, and
we adopt units for which $\hbar = c = 1$.  For late times the
growing solution to this equation has the form
\beq
  a(t) \propto e^{\chi t} \ , \hbox{ where } \chi = \sqrt{{8 \pi
   \over 3} G \rho_f } \ .
  \label{eq:2.6}
\eeq
Of course inflationary theorists prefer not to assume that the
universe began homogeneously and isotropically, but there is
considerable evidence for the ``cosmological no-hair conjecture''
\cite{Jensen-Stein-Schabes}, which implies that a wide class of
initial states will approach this exponentially expanding
solution. 

So the basic scenario of new inflation begins by assuming
that at least some patch of the early universe was in this
peculiar false vacuum state.  In the original papers
\cite{Linde1, Albrecht-Steinhardt1} this initial
condition was motivated by the fact that, in many quantum field
theories, the false vacuum resulted naturally from the
supercooling of an initially hot state in thermal equilibrium. 
It was soon found, however, that quantum fluctuations in the
rolling inflaton field give rise to density perturbations in the
universe \cite{Starobinsky2, GuthPi, Hawking1, BST, BFM}, and
that these density perturbations would be much larger than
observed unless the inflaton field is very weakly coupled.  For
such weak coupling there would be no time for an initially
nonthermal state to reach thermal equilibrium.  Nonetheless,
since thermal equilibrium describes a probability distribution in
which all states of a given energy are weighted equally, the fact
that thermal equilibrium leads to a false vacuum implies that
false vacuum-like states are not uncommon.  Thus, even in the
absence of thermal equilibrium, even if the universe started in a
highly chaotic initial state, it seems reasonable to simply
assume that some small patches of the early universe settled into
the false vacuum state, as was suggested for example in
\cite{Guth-RS}.  The idea that one should consider small
patches of the early universe with arbitrary initial
configurations of scalar fields was later emphasized by Linde
\cite{chaotic} in the context of chaotic inflation.  Linde
pointed out that even highly improbable initial patches could be
important if they inflated, since the exponential expansion could
still cause such patches to dominate the volume of the universe. 
One might hope that eventually a full theory of quantum origins
would allow us to calculate the probability of regions settling
into the false vacuum, but I will argue in Sec.~V that, in the
context of eternal inflation, this probability is quite
irrelevant.
 
Once a region of false vacuum materializes, the physics of the
subsequent evolution seems rather clear-cut.  The gravitational
repulsion caused by the negative pressure will drive the region
into a period of exponential expansion.  If the energy density of
the false vacuum is at the grand unified theory scale ($\rho_f
\approx (2 \times 10^{16}\ \hbox{GeV})^4)$, Eq.~(\ref{eq:2.6})
shows that the time constant $\chi^{-1}$ of the exponential
expansion would be about $10^{-38}$ sec.  For inflation to
achieve its goals, this patch has to expand exponentially for at
least 60 e-foldings.  Then, because this state is only
metastable---the inflaton field is perched on top of the hill of
the potential energy diagram of Fig.~\ref{newinf}---eventually
this state will decay.  The inflaton field will roll off the
hill, ending inflation.  And when it does, the energy density
that has been locked in the inflaton field is released. Because
of the coupling of the inflaton to other fields, that energy
becomes thermalized to produce a hot soup of particles, which is
exactly what had always been taken as the starting point of the
standard big bang theory before inflation was introduced.  From
here on the scenario joins onto the standard big bang
description.  The role of inflation is to replace the postulates
of the standard big bang theory with dynamically generated
initial conditions.

The inflationary mechanism produces an entire universe starting
from essentially nothing, so one needs to answer the question of
where the energy of the universe came from.  The answer is that
it came from the gravitational field.  I am not saying that the
colossal energy of the universe was stored from the beginning in
the gravitational field.  Rather, the crucial point is that the energy
density of the gravitational field is literally negative---a
statement which is true both in Newtonian gravity and in general
relativity.  So, as more and more positive energy  materialized
in the form of an ever-growing region filled with a
high-energy scalar field, more and more negative energy 
materialized in the form of an expanding region filled with a
gravitational field.  So the total energy remained very small, and
could in fact be exactly zero.  There is nothing known that
places any limit on the amount of inflation that can occur while
the total energy remains exactly zero.\footnote{In Newtonian
mechanics the energy density of a gravitational field is
unambiguously negative; it can be derived by the same methods
used for the Coulomb field, but the force law has the opposite
sign.  In general relativity there is no coordinate-invariant way
of expressing the energy in a space that is not asymptotically
flat, so many experts prefer to say that the total energy is
undefined.  Either way, there is agreement that inflation is
consistent with the general relativistic description of energy
conservation.}

\subhead{Chaotic Inflation:}

\begin{figure}[ht]
\epsfxsize=275pt 
\centerline{\epsfbox{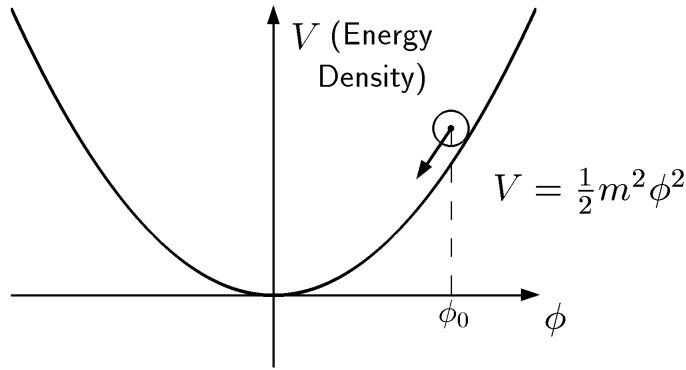}}
\caption{Generic form of the potential for the chaotic inflationary
scenario.} 
\label{chaoticinf}
\end{figure}

Chaotic inflation \cite{chaotic} can occur in the context of a
much more general class of potential energy functions.  In
particular, even a potential energy function as simple as
Eq.~(\ref{eq:2.1}), describing a scalar field with a mass and no
interaction, is sufficient to describe chaotic inflation. 
Chaotic inflation is illustrated in Fig.~\ref{chaoticinf}.  In
this case there is no state that bears any obvious resemblance to
the false vacuum of new inflation.  Instead the scenario works by
supposing that chaotic conditions in the early universe produced
one or more patches in which the inflaton field $\phi$ was at
some high value $\phi = \phi_0$ on the potential energy curve. 
Inflation occurs as the inflaton field rolls down the hill.  As
long as the initial value $\phi_0$ is sufficiently high on the
curve, there will be sufficient inflation to solve all the
problems that inflation is intended to solve. 

The equations describing chaotic inflation can be written simply,
provided that we assume that the universe is already flat enough
so that we do not need to include a curvature term.  The field
equation for the inflaton field in the expanding universe is
\beq
  \ddot \phi + 3 H \dot \phi = - {\partial V \over \partial
          \phi } \ ,
  \label{eq:2.7}
\eeq
where the overdot denotes a derivative with respect to time $t$,
and $H$ is the time-dependent Hubble parameter given by
\beq
  H^2 = {8 \pi \over 3} G V  \ .
  \label{eq:2.8}
\eeq
For the toy-model potential energy of Eq.~(\ref{eq:2.1}), these
equations have a very simple solution:
\beq
  \phi = \phi_0 - {m \over \sqrt{12 \pi G}} \, t \ .
  \label{eq:2.9}
\eeq
One can then calculate the number $N$ of inflationary e-foldings,
which is given by
\beq
  N = \int_{\phi = \phi_0}^{\phi = 0} H(t) \, dt = 2 \pi G
     \phi_0^2 \ .
  \label{eq:2.10}
\eeq
In this free-field model $N$ depends only on $\phi_0$ and not on the
inflaton mass $m$.  Thus the number of e-foldings will exceed 60
provided that
\beq
  \phi_0 > \sqrt{60 \over 2 \pi} \, M_{\rm P} \approx 3.1 M_{\rm
  P} \ , \label{eq:2.11}
\eeq
where $M_{\rm P} \equiv 1/\sqrt{G} = 1.22 \times 10^{19}$ GeV is
the Planck mass.  Although the value of the scalar field is
larger than $M_{\rm P}$, the energy density can be low compared
to the Planck scale: 
\beq
  \rho_0 = {1 \over 2} m^2 \phi_0^2 > {60 \over 4 \pi} M_{\rm P}^2 m^2
     \ .
  \label{eq:2.12}
\eeq
For example, if $m = 10^{16}$ GeV, then the potential energy
density is only $3 \times 10^{-6}\, M_{\rm P}^4$.  Since it is
presumably the energy density that is relevant to gravity, one
does not expect this situation to lead to strong quantum gravity
effects.

\section{Evidence for Inflation}
\setcounter{equation}{0}

The arguments in favor of inflation are pretty much the same no
matter which form of inflation we are discussing.  In my opinion,
the evidence that our universe is the result of some form of
inflation is very solid.  Since the term {\it inflation}
encompasses a wide range of detailed theories, it is hard to
imagine any alternative.  Let me review the basic arguments.

\subhead{1) The universe is big}

First of all, we know that the universe is incredibly large.  The
visible part of it contains about $10^{90}$ particles.  It is
easy, however, to take this fact for granted: of course the
universe is big, it's the whole universe!  In ``standard''
Friedmann-Robertson-Walker cosmology, without inflation, one
simply postulates that about $10^{90}$ or more particles were
here from the start.  If, however, we try to imagine a theory
describing the origin of the universe, it would have to somehow
output this number of $10^{90}$ or more.  That is a very big
number, and it is hard to imagine it ever coming out of a
calculation in which the input consists only geometrical
quantities, quantities associated with simple dynamics, and
factors of 2 and $\pi$.  In the inflationary model, the huge
number of particles is explained naturally by the exponential
expansion, which reduces the problem to explaining 60 or 70
e-foldings of inflation.  In fact, it is easy to construct
underlying particle theories that will give far more than 70
e-foldings, suggesting that the observed universe is only a tiny
speck within the universe as a whole.

\subhead{2) The Hubble expansion}

The Hubble expansion is also easy to take for granted, since it
is so familiar.  In standard FRW cosmology, the Hubble expansion
is part of the list of postulates that define the initial
conditions.  But inflation offers an explanation of how the
Hubble expansion began.  The repulsive gravity associated with
the false vacuum is exactly the kind of force needed to propel
the universe into a pattern of motion in which any two particles
are moving apart with a velocity proportional to their
separation.

\subhead{3) Homogeneity and isotropy}

The degree of uniformity in the universe is startling.  Through
careful measurements of the cosmic background radiation, we know
that the intensity of this radiation is the same in all
directions to an accuracy of 1 part in 100,000.  To get some
feeling for how high this precision is, we can imagine a marble
that is spherical to this accuracy.  The surface of the
marble would have to be shaped with a tolerance of about 1,000
angstroms, a quarter of the wavelength of light. 

Although precision lenses can be ground to quarter-wavelength
accuracy, we would nonetheless be shocked if we ever dug up a
stone from the ground that was round to this extraordinary
accuracy.  If such a stone were somehow found, I am confident
that we would not accept an explanation of its origin which
simply proposed that the stone started out perfectly round. 
Similarly, in the current era, I do not think it makes sense to
consider any theory of cosmogenesis that cannot offer some
explanation of how the universe became so incredibly isotropic. 

The uniformity of the cosmic background radiation implies that
the observed universe had become uniform in temperature by about
300,000 years after the big bang, when the universe cooled enough
so that the opaque plasma neutralized into a transparent gas.  In
standard FRW cosmology, the uniformity could be established by
this time only if signals could propagate 100 times faster than
light, which is not possible.  In inflationary cosmology,
however, the uniformity can be created initially on microscopic
scales, by normal thermal-equilibrium processes.  Then inflation
takes over and stretches the regions of uniformity to become
large enough to encompass the observed universe.

\subhead{4) The flatness problem}

I find the flatness problem particularly impressive, because the
numbers that it leads to are so extraordinary.  The problem
concerns the value of the ratio
\beq
  \Omega_\tot \equiv {\rho_\tot \over \rho_c} \ ,
  \label{eq:3.1}
\eeq
where $\rho_\tot$ is the average total mass density of the
universe and $\rho_c = 3 H^2 / 8 \pi G$ is the critical density,
the density that would make the universe spatially flat. 
($\rho_\tot$ includes any vacuum energy $\rho_{\rm vac} =
\Lambda/ 8 \pi G$ associated with the cosmological constant
$\Lambda$, if it is nonzero.)

The present value of $\Omega_\tot$ satisfies
\beq
  0.1 \lta \Omega_{\tot,0} \lta 2 \ ,
  \label{eq:3.2}
\eeq
but the precise value is not known.  Despite the breadth of this
range, the value of $\Omega_\tot$ at early times is highly
constrained, since $\Omega_\tot=1$ is an unstable equilibrium
point of the standard model evolution.  If $\Omega_\tot$ was ever
{\it exactly} equal to one, it would remain so forever.  However,
if $\Omega_\tot$ differed slightly from 1 in the early universe,
that difference---whether positive or negative---would be
amplified with time.  In particular, the FRW equations imply that
$\Omega_\tot - 1$ grows as
\beq
  \Omega_\tot - 1 \propto \cases{t &(during the
     radiation-dominated era)\cr 
    t^{2/3} &(during the matter-dominated era)\ .\cr}
  \label{eq:3.3}
\eeq
At $t=1$ sec, for example, Dicke and Peebles \cite{dicke} pointed
out that $\Omega_\tot$ must have equaled one to an accuracy of
one part in $10^{15}$.  Classical cosmology provides no
explanation for this fact---it is simply assumed as part of the
initial conditions.  In the context of modern particle theory,
where we try to push things all the way back to the Planck time,
$10^{-43}$ sec, the problem becomes even more extreme.  At this
time $\Omega_\tot$ must have equaled one to 58 decimal places!

While this extraordinary flatness of the early universe has no
explanation in classical FRW cosmology, it is a natural
prediction for inflationary cosmology.  During the inflationary
period, instead of $\Omega_\tot$ being driven away from 1 as
described by Eq.~(\ref{eq:3.3}), $\Omega_\tot$ is driven towards
1, with exponential swiftness:
\beq
  \Omega_\tot - 1 \propto e^{-2 H_{\rm inf} t} \ ,
  \label{eq:3.4}
\eeq
where $H_{\rm inf}$ is the Hubble parameter during inflation. 
Thus, as long as there is enough inflation, $\Omega_\tot$ can
start at almost any value, and it will be driven to unity by the
exponential expansion. 

\subhead{5) Absence of magnetic monopoles}

All grand unified theories predict that there should be, in the
spectrum of possible particles, extremely massive magnetic
monopoles.  By combining grand unified theories
with classical cosmology without inflation, Preskill
\cite{preskill} found that magnetic monopoles would be produced
so copiously that they would outweigh everything else in the
universe by a factor of about $10^{12}$.  A mass density this
large would cause the inferred age of the universe to drop to
about 30,000 years!  In inflationary models, the monopoles can be
eliminated simply by arranging the parameters so that inflation
takes place after (or during) monopole production, so the
monopole density is diluted to a completely negligible level.

\subhead{6) Anisotropy of the cosmic background radiation}

The process of inflation smooths the universe essentially
completely, but quantum fluctuations of the inflaton field can
generate density fluctuations as inflation ends.  Generically
these are adiabatic Gaussian fluctuations with a nearly
scale-invariant spectrum \cite{Starobinsky2, GuthPi, Hawking1,
BST, BFM}.  New data is arriving quickly, but so far the
observations are in excellent agreement with the predictions of
the simplest inflationary models.  For a review, see for example
Bond and Jaffe \cite{bond-jaffe}, who find that the combined data
give a slope of the primordial power spectrum within 5\% of the
preferred scale-invariant value.

\section{Eternal Inflation: Mechanisms}
\setcounter{equation}{0}

Having discussed the mechanisms and the motivation for inflation
itself, I now wish to move on the main issue that I want to
stress in this article---eternal inflation, the questions that it
can answer, and the questions that it raises.  In this section I
will discuss the mechanisms that make eternal inflation possible,
leaving the other issues for the following sections.  I will
discuss eternal inflation first in the context of new inflation,
and then in the context of chaotic inflation, where it is more
subtle. 

\subhead{Eternal New Inflation:}

The eternal nature of new inflation was first discovered by
Steinhardt \cite{steinhardt-nuffield} and Vilenkin
\cite{vilenkin-eternal} in 1983.  Although the false vacuum is a
metastable state, the decay of the false vacuum is an exponential
process, very much like the decay of any radioactive or unstable
substance.  The probability of finding the inflaton field at the
top of the plateau in its potential energy diagram does not fall
sharply to zero, but instead trails off exponentially with time
\cite{guth-pi2}.  However, unlike a normal radioactive substance
such as radium, the false vacuum exponentially expands at the
same time that it decays. In fact, in any successful inflationary
model the rate of exponential expansion is always much faster
than the rate of exponential decay.  Therefore, even though the
false vacuum is decaying, it never disappears, and in fact the
total volume of the false vacuum, once inflation starts,
continues to grow exponentially with time, ad infinitum. 

\begin{figure}[ht]
\centerline{\epsfbox{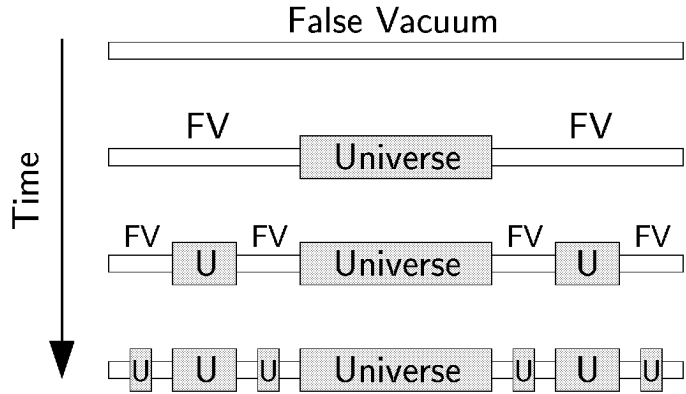}}
\caption{A schematic illustration of eternal inflation.} 
\label{eternalline}
\end{figure}

Fig.~\ref{eternalline} shows a schematic diagram of an eternally
inflating universe.  The top bar indicates a region of false
vacuum.  The evolution of this region is shown by the successive
bars moving downward, except that I could not show the expansion
and still fit all the bars on the page.  So the region is shown
as having a fixed size in comoving coordinates, while the scale
factor, which is not shown, increases from each bar to the next. 
As a concrete example, suppose that the scale factor for each bar
is three times larger than for the previous bar.  If we follow
the region of false vacuum indicated by the top bar as it evolves
into the second bar, in about one third of the region the scalar
field rolls down the hill of the potential energy diagram,
precipitating a local big bang that will evolve into something
that will eventually appear to its inhabitants as a universe. 
This local big bang region is shown in gray and labeled
``Universe.''  Meanwhile, however, the space has expanded so much
that each of the two remaining regions of false vacuum is the
same size as the starting region.  Thus, if we follow the region
for another time interval of the same duration, each of these
regions of false vacuum will break up, with about one third of
each evolving into a local universe, as shown on the third bar
from the top.  Now there are four remaining regions of false
vacuum, and again each is as large as the starting region.  This
process will repeat itself literally forever, producing a kind of
a fractal structure to the universe, resulting in an infinite
number of the local universes shown in gray.  These local
universes are often called {\it bubble universes,} but that
terminology conveys the unfortunate connotation that the local
universes are spherical.  While bubbles formed in first-order
phase transitions are round \cite{coleman-deluccia}, the local
universes formed in eternal new inflation are generally very
irregular, as can be seen for example in the two-dimensional
simulation by Vanchurin, Vilenkin, and Winitzki in Fig.~2 of
Ref.~\cite{vvw}.  I therefore prefer to call them {\it pocket
universes,} to try to avoid the suggestion that they are round.

The diagram in Fig.~\ref{eternalline} is of course an
idealization.  The real universe is three dimensional, while the
diagram illustrates a schematic one-dimensional universe.  It is
also important that the decay of the false vacuum is really a
random process, while I constructed the diagram to show a very
systematic decay, because it is easier to draw and to think
about.  When these inaccuracies are corrected, we are still left
with a scenario in which inflation leads asymptotically to a
fractal structure \cite{aryal-vilenkin} in which the universe as
a whole is populated by pocket universes on arbitrarily small
comoving scales.  Of course this fractal structure is entirely on
distance scales much too large to be observed, so we cannot
expect astronomers to actually find it.  Nonetheless, one does
have to think about the fractal structure if one wants to
understand the very large scale structure of the spacetime
produced by inflation. 

Most important of all is the simple statement that once inflation
begins, it produces not just one universe, but an infinite
number of universes. 

\subhead{Eternal Chaotic Inflation:}

The eternal nature of new inflation depends crucially on the
scalar field lingering at the top of the plateau of
Fig.~\ref{newinf}.  Since the potential function for chaotic
inflation, Fig.~\ref{chaoticinf}, has no plateau, it does not
seem likely that eternal inflation can happen in this context. 
Nonetheless, Andrei Linde \cite{linde-eternal} showed in 1986
that chaotic inflation can also be eternal. 

The key to eternal chaotic inflation is the role of quantum
fluctuations, which is very significant in all inflationary
models.  Quantum fluctuations are invariably important on very
small scales, and with inflation these very small scales are
rapidly stretched to become macroscopic and even astronomical. 
Thus the quantum fluctuations of the inflaton field can have very
noticeable effects.

\begin{figure}[ht]
\epsfxsize=275pt 
\centerline{\epsfbox{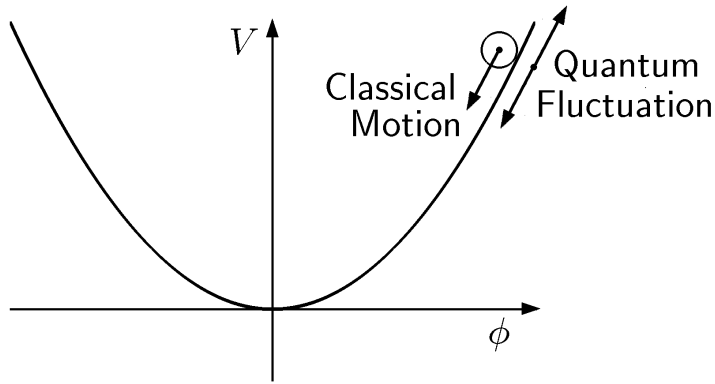}}
\caption{Evolution of the inflaton field during eternal chaotic
inflation.} 
\label{chaotic-eternal}
\end{figure}

When the mass of the scalar field is small compared to the Hubble
parameter $H$, these quantum evolution of the scalar field is
accurately described as a random walk.  It is useful to divide
space into regions of physical size $H^{-1}$, and to discuss the
average value of the scalar field $\phi$ within a given region. 
In a time $H^{-1}$, the quantum fluctuations cause the scalar
field to undergo a random Gaussian jump of zero mean and a
root-mean-squared magnitude
\cite{random-vil-ford,random-linde,Starobinsky2,random-starobinsky}
given by
\beq
  \Delta \phi_{\rm qu} = {H \over 2 \pi} \ .
  \label{eq:4.1}
\eeq
This random quantum jump is superimposed on the classical motion,
as indicated in Fig.~(\ref{chaotic-eternal}).

To illustrate how eternal inflation happens in the simplest
context, let us consider again the free scalar field described by
the potential function of Eq.~(\ref{eq:2.1}).  We consider a
region of physical radius $H^{-1}$, in which the field has an
average value $\phi$.  Using Eq.~(\ref{eq:2.9}) along with
Eqs.~(\ref{eq:2.8}) and (\ref{eq:2.1}), one finds that the
magnitude of the classical change that the field will undergo in
a time $H^{-1}$ is given in by
\beq
  \Delta \phi_{\rm cl} = {M_{\rm P} m \over \sqrt{12 \pi}} \, H^{-1} =
     {1 \over 4 \pi} {M_{\rm P}^2 \over \phi} \ .
  \label{eq:4.2}
\eeq
Let $\phi^*$ denote the value of $\phi$ which is
large enough so that
\beq
  \Delta \phi_{\rm qu} (\phi^*) = \Delta \phi_{\rm cl}(\phi^*) \ ,
  \label{eq:4.3}
\eeq
which implies that
\beq
  \phi^* = \left( {3 \over 16 \pi} \right)^{1/4} {M_{\rm P}^{3/2} \over
     m^{1/2}} \ .
  \label{eq:4.4}
\eeq

Now consider what happens to a region for which the initial
average value of $\phi$ is equal to $\phi^*$.  In a time interval
$H^{-1}$, the volume of the region will increase by $e^3 \approx
20$.  At the end of the time interval we can divide the original
region into 20 regions of the same volume as the original, and in
each region the average scalar field can be written as
\beq
  \phi = \phi^* + \Delta \phi_{\rm cl} + \delta \phi \ ,
  \label{eq:4.5}
\eeq
where $\delta \phi$ denotes the random quantum jump, which is
drawn from a Gaussian probability distribution with standard
deviation $\Delta \phi_{\rm qu} = \Delta \phi_{\rm cl}$. 
Gaussian statistics imply that there is a 15.9\% chance that a
Gaussian random variable will exceed its mean by more than one
standard deviation, and therefore there is a 15.9\% chance that
the net change in $\phi$ will be positive.  Since there are now
20 regions of the original volume, on average the value of $\phi$
will exceed the original value in 3.2 of these regions.  Thus the
volume for which $\phi \ge \phi^*$ does not (on average)
decrease, but instead increases by more than a factor of 3. 
Since this argument can be repeated, the expectation value of the
volume for which $\phi \ge \phi^*$ increases exponentially with
time.  Typically, therefore, inflation never ends, but instead
the volume of the inflating region grows exponentially without
bound.  The minimum field value for eternal inflation is a little
below $\phi^*$, since a volume increase by a factor of 3.2 is
more than necessary---any factor greater than one would be
sufficient.  A short calculation shows that the minimal value for
eternal inflation is $0.78 \phi^*$.

While the value of $\phi^*$ is larger than Planck scale, again we
find that this is not true of the energy density:
\beq
  V(\phi^*) = {1 \over 2} m^2 \phi^{*2} = \sqrt{3 \over 64 \pi}
     \, m M_{\rm P}^3 \ ,
  \label{eq:4.6}
\eeq
which for $m = 10^{16}$ GeV gives an energy density of $1 \times
10^{-4} \, M_{\rm P}^4$.  

If one carries out the same analysis with a potential function
\beq
  V(\phi) = {1 \over 4} \lambda \phi^4 \ ,
  \label{eq:4.7}
\eeq
one finds \cite{linde-book} that
\beq
  \phi^* = \left( 3 \over 2 \pi \lambda \right)^{1/6} M_{\rm P} \ ,
  \label{eq:4.8}
\eeq
and
\beq
  V(\phi^*) = \left( 3 \over 16 \pi \right)^{2/3} \lambda^{1/3}
     M_{\rm P}^4 \ .
  \label{eq:4.9}
\eeq
Since  $\lambda$ must be very small in any case so that
density perturbations are not too large, one finds again that
eternal inflation is predicted to happen at an energy density well
below the Planck scale.

\section{Eternal Inflation: Implications}
\setcounter{equation}{0}
\label{implications}

When I told Rocky Kolb that I was going to be talking about
eternal inflation, he said, ``That's OK, we can talk about
physics later.''  So that's the point I'd like to address here. 
In spite of the fact that the other universes created by eternal
inflation are too remote to imagine observing directly, I still
believe that eternal inflation has real consequences in terms of
the way we extract predictions from theoretical models. 
Specifically, there are four consequences of eternal inflation
that I will highlight.
 
\subhead{1) Unobservability of initial conditions}

First, eternal inflation implies that all hypotheses about the
ultimate initial conditions for the universe---such as the
Hartle-Hawking \cite{hartle-hawking} no boundary proposal, the
tunneling proposals by Vilenkin \cite{tunnel-vilenkin} or Linde
\cite{tunnel-linde}, or the more recent Hawking-Turok instanton
\cite{hawking-turok}---become totally divorced from observation.
That is, one would expect that if inflation is to continue
arbitrarily far into the future with the production of an
infinite number of pocket universes, then the statistical
properties of the inflating region should approach a steady state
which is independent of the initial conditions.  Unfortunately,
attempts to quantitatively study this steady state are severely
limited by several factors.  First, there are ambiguities in
defining probabilities, which will be discussed later.  In
addition, the steady state properties seem to depend strongly on
super-Planckian physics which we do not understand.  That is, the
same quantum fluctuations that make eternal chaotic inflation
possible tend to drive the scalar field further and further up
the potential energy curve, so attempts to quantify the steady
state probability distribution \cite{LLM,GBLinde} require the
imposition of some kind of a boundary condition at large $\phi$. 
Although these problems remain unsolved, I still believe that it
is reasonable to assume that in the course of its perpetual
evolution, an eternally inflating universe would lose all memory
of the state in which it started.

Although the ultimate origin of the universe would become
unobservable, I would not expect that the question of how the
universe began would lose its interest.  While eternally
inflating universes continue forever once they start, they are
presumably not eternal into the past.  (The word {\it eternal} is
therefore not technically correct---it would be more precise to
call this scenario {\it semi-eternal} or {\it future-eternal}.) 
While the issue is not completely settled, it appears likely that
eternally inflating universes must necessarily have a beginning. 
Borde and Vilenkin \cite{borde-vilenkin} have shown, provided
that certain conditions are met, that spacetimes which are
future-eternal must have an initial singularity, in the sense
that they cannot be past null geodesically complete.  The proof,
however, requires the weak energy condition, which is classically
valid but quantum-mechanically violated \cite{borde-vilenkin2}. 
In any case, I am not aware of any viable model without a
beginning, and certainly nothing that we know can rule out the
possibility of a beginning.  The possibility of a quantum origin
of the universe is very attractive, and will no doubt be a
subject of interest for some time.  Eternal inflation, however,
seems to imply that the entire study will have to be conducted
with literally no input from observation. 

\subhead{2) Irrelevance of initial probability}

A second consequence of eternal inflation is that the
probability of the onset of inflation becomes totally
irrelevant, provided that the probability is not identically zero.
Various authors in the past have argued that one type of
inflation is more plausible than another, because the initial
conditions that it requires appear more likely to have occurred. 
In the context of eternal inflation, however, such arguments have
no significance.

To illustrate the insignificance of the probability of the onset
of inflation, I will use a numerical example.  We will imagine
comparing two different versions of inflation, which I will call
Type A and Type B\relax.  They are both eternally inflating---but
Type A will have a higher probability of starting, while Type B
will be a little faster in its exponential expansion rate.  Since
I am trying to show that the higher starting probability of Type
A is irrelevant, I will choose my numbers to be extremely
generous to Type A\relax.  First, we must choose a number for how
much more probable it is for Type A inflation to begin, relative
to type B\relax.  A googol, $10^{100}$, is usually considered a
large number---it is some 20 orders of magnitude larger than the
total number of baryons in the visible universe.  But I will be
more generous: I will assume that Type A inflation is more likely
to start than type B inflation by a factor of $10^{1,000,000}$. 
Type B inflation, however, expands just a little bit faster, say
by 0.001\%.  We need to choose a time constant for the
exponential expansion, which I will take to be a typical grand
unified theory scale, $\tau = 10^{-37}$ sec. ($\tau$ represents
the time constant for the overall expansion factor, which takes
into account both the inflationary expansion and the exponential
decay of the false vacuum.)  Finally, we need to choose a length
of time to let the system evolve.  In principle this time
interval is infinite (the inflation is eternal into the future),
but to be conservative we will watch the system for only one
second.

We imagine setting up a statistical ensemble of universes at
$t=0$, with an expectation value for the volume of Type A
inflation exceeding that of Type B inflation by $10^{1,000,000}$. 
For brevity, let the term ``weight'' to refer to the ensemble
expectation value of the volume.  Thus, the weights of Type A
inflation and Type B inflation will begin with the ratio
\beq
  \left.{W_B \over W_A}\right|_{t=0} = 10^{-1,000,000} \ .
  \label{eq:5.1}
\eeq
After one second of evolution, the expansion factors for Type A
and Type B inflation will be
\begin{eqnarray}
  \label{eq:5.2}
  Z_A & = & e^{t/\tau} = e^{10^{37}} \\
  \label{eq:5.3}
  Z_B & = & e^{1.00001\,t/\tau} = e^{0.00001\,t/\tau} Z_A
          \nonumber \\
      & = & e^{10^{32}} Z_A \approx 10^{4.3 \times 10^{31}} Z_A 
          \ .
\end{eqnarray}
The weights at the end of one second are proportional to these
expansion factors, so
\beq
  \left.{W_B \over W_A}\right|_{t=1\ \rm sec} = 10^{\left(4.3
     \times 10^{31} - 1,000,000\right)} \ .
  \label{eq:5.4}
\eeq
Thus, the initial ratio of $10^{1,000,000}$ is vastly superseded
by the difference in exponential expansion factors.  In fact, we
would have to calculate the exponent of Eq.~(\ref{eq:5.4}) to an
accuracy of 25 significant figures to be able to barely detect
the effect of the initial factor of $10^{1,000,000}$. 

One might criticize the above argument for being naive, as the
concept of time was invoked without any discussion of how the
equal-time hypersurfaces are to be chosen.  I do not know a
decisive answer to this objection; as I will discuss later, there
are unresolved questions concerning the calculation of
probabilities in eternally inflating spacetimes.  Nonetheless,
given that there is actually an infinity of time available, it is
seems reasonable to believe that the form of inflation that
expands the fastest will always dominate over the slower forms by
an infinite factor.

A corollary to this argument is that new inflation is not dead. 
While the initial conditions necessary for new inflation cannot
be justified on the basis of thermal equilibrium, as proposed in
the original papers \cite{Linde1,Albrecht-Steinhardt1}, in the
context of eternal inflation it is sufficient to conclude that
the probability for the required initial conditions is nonzero. 
Since the resulting scenario does not depend on the words that
are used to justify the initial state, the standard treatment of
new inflation remains valid.

\subhead{3) Inevitability of eternal inflation}

Third, I'd like to claim that, since it appears that a universe
is in principle capable of eternally reproducing, it is hard to
believe that any other description can make sense at all.  To
clarify this point, let me raise the analogy of rabbits.  We all
know that rabbits can reproduce---in fact, they reproduce like
rabbits.  Suppose that you went out into the woods and found a
rabbit that had characteristics indicating that it did not belong
to any known rabbit species.  Then you would have to theorize
about how the rabbit originated.  You might entertain the notion
that the rabbit was created by some unique, mysterious, cosmic
event that you hope to someday understand better.  Or you could
assume that the rabbit was created by the process of rabbit
reproduction that we all know so well.  I think that we would all
consider the latter possibility to be far more plausible.  So, I
claim that once we become convinced that universes can reproduce
like rabbits, then the situations are similar.  When we notice
that there is a universe and ask how it originated, the same
inferences that we made for the rabbit question should apply to
this one.

\subhead{4) Possibility of restoring the uniqueness of
theoretical predictions}

A fourth consequence of eternal inflation is the possibility that
it offers to rescue the predictive power of theoretical physics. 
All the indications suggest that string theory or M theory
describes an elegantly unique theoretical structure, but
nonetheless it seem unlikely that the theory possesses a unique
vacuum.  Since predictions will ultimately depend on the
properties of the vacuum, the predictive power of string/M theory
may be limited.  Eternal inflation, however, provides a hope that
this problem can be remedied.  Even if many types of vacua are
equally stable, it may turn out that a unique state produces the
maximum possible rate of inflation.  If so, then this state will
dominate the universe, even if its expansion rate is only
infinitesimally larger than the other possibilities.  Thus,
eternal inflation might allow physicists to extract unique
predictions, in spite of the multiplicity of stable vacua.

\section{Difficulties in Calculating Probabilities}
\setcounter{equation}{0}

In an eternally inflating universe, anything that can happen will
happen; in fact, it will happen an infinite number of times.  Thus,
the question of what is possible becomes trivial---anything is
possible, unless it violates some absolute conservation law.  To
extract predictions from the theory, we must therefore learn to
distinguish the probable from the improbable.

However, as soon as one attempts to define probabilities in an
eternally inflating spacetime, one discovers ambiguities.  Since
an eternally inflating universe produces an infinite number of
pocket universes, the sample space is infinite.  The fraction of
universes with any particular property is given by the
meaningless ratio of infinity divided by infinity.  To obtain a
well-defined answer, one needs to invoke some method of
regularization.  The most straightforward form of regularization
consists of truncating the space to a finite subspace, and then
taking a limit in which the subspace becomes larger and larger.

To understand the nature of the problem, it is useful to think
about the integers as a model system with an infinite number of
entities.  We can ask, for example, what fraction of the integers
are odd.  Most people would presumably say that the answer is
$1/2$, since the integers alternate between odd and even.  That
is, if the string of integers is truncated after the $N$th, then
the fraction of odd integers in the string is exactly $1/2$ if
$N$ is even, and is $(N+1)/2N$ if $N$ is odd.  In any case, the
fraction approaches $1/2$ as $N$ approaches infinity.

However, the ambiguity of the answer can be seen if one imagines
other orderings for the integers.  One could, if one wished,
order the integers as 
\beq
  1,3,\ 2,\ 5,7,\ 4,\ 9,11,\ 6\ ,\ldots,  
  \label{eq:6.1}
\eeq
always writing two odd integers followed by one even integer. 
This series includes each integer exactly once, just like the
usual sequence ($1,2,3,4, \ldots$).  The integers are just
arranged in an unusual order.  However, if we truncate the
sequence shown in Eq.~(\ref{eq:6.1}) after the $N$th entry, and
then take the limit $N \to \infty$, we would conclude that 2/3 of
the integers are odd.  Thus, we see that probabilities can
depend nontrivially on the method of regularization that is used.

In the case of eternally inflating spacetimes, one might consider
a regularization defined by ordering the pocket universes in the
sequence in which they form, and then truncating after the $N$th. 
However, each pocket universe fills its own future light cone, so
no pocket universe forms in the future light cone of another. 
Any two pocket universes are spacelike separated from each other,
so different observers can disagree about which formed first. 
One can arbitrarily choose equal-time surfaces that foliate the
spacetime, and then truncate at some value of $t$, but this
recipe is far from unique.  In practice, different ways of
choosing equal-time surfaces give different results. 

\section{The Youngness Paradox}
\setcounter{equation}{0}

If one chooses a regularization in the most naive way, one is led
to a set of very peculiar results which I call the {\it youngness
paradox.} 

Specifically, suppose that one constructs a Robertson-Walker
coordinate system while the model universe is still in the false
vacuum (de Sitter) phase, before any pocket universes have
formed. One can then propagate this coordinate system forward
with a synchronous gauge condition,\footnote{By a synchronous
gauge condition, I mean that each equal-time hypersurface is
obtained by propagating every point on the previous hypersurface
by a fixed infinitesimal time interval $\Delta t$ in the
direction normal to the hypersurface.} and one can define
probabilities by truncating at a fixed value $t_f$ of the
synchronous time coordinate $t$.  That is, the probability of any
particular property can be taken to be proportional to the volume
on the $t = t_f$ hypersurface which has that property.  This
method of defining probabilities was studied in detail by Linde,
Linde, and Mezhlumian, in a paper with the memorable title ``Do
we live in the center of the world?'' \cite{center-world}.  I
will refer to probabilities defined in this way as synchronous
gauge probabilities.

The youngness paradox is caused by the fact that the volume of
false vacuum is growing exponentially with time with an
extraordinarily short time constant, in the vicinity of
$10^{-37}$ sec.  Since the rate at which pocket universes form is
proportional to the volume of false vacuum, this rate is
increasing exponentially with the same time constant.  That means
that in each second the number of pocket universes that exist is
multiplied by a factor of $\exp\left\{10^{37}\right\}$.  At any
given time, therefore, almost all of the pocket universes that
exist are universes that formed very very recently, within the
last several time constants.  The population of pocket universes
is therefore an incredibly youth-dominated society, in which the
mature universes are vastly outnumbered by universes that have
just barely begun to evolve.  Although a mature universe has a
larger volume then a young one, this multiplicative factor is of
little importance, since in synchronous coordinates the volume no
longer grows exponentially once the pocket universe forms.

Probability calculations in this youth-dominated ensemble lead to
peculiar results, as discussed in Ref.~\cite{center-world}.  These
authors considered the expected behavior of the mass density in
our vicinity, concluding that we should find ourselves very near
the center of a spherical low-density region.  Here I would like
to discuss a less physical but simpler question, just to
illustrate the paradoxes associated with synchronous gauge
probabilities.  Specifically, I will consider the question: ``Are
there any other civilizations in the visible universe that are
more advanced than ours?''.  Intuitively I would not expect
inflation to make any predictions about this question, but I will
argue that the synchronous gauge probability distribution
strongly implies that there is no civilization in the visible
universe more advanced than we are.

Suppose that we have reached some level of advancement, and
suppose that $t_{\rm min}$ represents the minimum amount of time
needed for a civilization as advanced as we are to evolve,
starting from the moment of the decay of the false vacuum---the
start of the big bang.  The reader might object on the grounds
that there are many possible measures of advancement, but I would
respond by inviting the reader to pick any measure she chooses;
the argument that I am about to give should apply to all of them. 
The reader might alternatively claim that there is no sharp
minimum $t_{\rm min}$, but instead we should describe the problem
in terms of a function which gives the probability that, for any
given region within a pocket universe of the size of our visible
universe, a civilization as advanced as we are would develop by
time $t$.  I believe, however, that the introduction of such a
probability distribution would merely complicate the argument,
without changing the result. So, for simplicity of discussion, I
will assume that there is some sharply defined minimum time
$t_{\rm min}$ required for a civilization as advanced as ours to
develop.

Since we exist, our pocket universe must have an age $t_0$
satisfying 
\beq
  t_0 \ge t_{\rm min} \ . 
  \label{eq:7.1}
\eeq
Suppose, however, that there is some civilization
in our visible universe that is more advanced than we are, let us
say by 1 second.  In that case Eq.~(\ref{eq:7.1}) is not
sufficient, but instead the age of our pocket universe would have
to satisfy
\beq
  t_0 \ge t_{\rm min} + 1 \hbox{\ second}\ . 
  \label{eq:7.2}
\eeq
However, in the synchronous gauge probability distribution,
universes that satisfy Eq.~(\ref{eq:7.2}) are outnumbered by
universes that satisfy Eq.~(\ref{eq:7.1}) by a factor of
approximately $\exp\left\{10^{37}\right\}$.  Thus, if we know
only that we are living in a pocket universe that satisfies
Eq.~(\ref{eq:7.1}), the probability that it also satisfies
Eq.~(\ref{eq:7.2}) is approximately
$\exp\left\{-10^{37}\right\}$.  We would conclude, therefore,
that it is extraordinarily improbable that there is a
civilization in our visible universe that is at least 1 second
more advanced than we are.

Perhaps this argument explains why SETI has not found any signals
from alien civilizations, but I find it more plausible that it is
merely a symptom that the synchronous gauge probability
distribution is not the right one.

\section{Toy Model of Eternal Inflation}
\setcounter{equation}{0}

The conceptual issue involved in the youngness paradox can
perhaps be clarified by considering a toy model of a highly
simplified eternally inflating universe.  Suppose that the
universe as a whole can be labeled with a global time variable
$t$, and that it consists of a countably infinite set of pocket
universes, each of which is labeled by an index $i$.  For
simplicity, we let each pocket universe have zero spatial
dimensions, so a spacetime point is fully specified by the time
$t$ and the index $i$ which indicates the pocket universe in
which it is located.  We assume that each pocket universe $i$
forms at some time $t_i = n_i \tau$, where $n_i$ is an integer and
$\tau$ is a fixed time constant characterizing the entire
universe.  Let the number of pocket universes that form at time
$t = n \tau$ be equal to $2^n$, for each nonnegative integer $n$. 
Assume that each pocket universe exists for a time $T \gg \tau$,
and then disappears, and that within each pocket universe the
interval from the time of formation to disappearance is uniformly
populated with ``sentient beings.''  Within each pocket universe
we can define a relative time, $t_{\rm rel} \equiv t - t_i$,
which measures the amount of time since the formation of the
pocket universe. 

The difficult question, then, is the following: At what relative
time $t_{\rm rel}$ does a {\it typical} sentient being live?  If
one answers this question by truncating the spacetime by the
criterion
\beq
  t \le t_c \ , 
  \label{eq:8.1}
\eeq
for some cut-off time $t_c = n_c \tau$, then one finds that most
of the pocket universes in the truncated spacetime formed within
the past few time constants.  As $t_c \to \infty$, the mean value
of $t_{\rm rel}$ approaches $\tau$.  This method is analogous to
the synchronous gauge cut-off discussed above.  If, however, one
truncates the spacetime by including all pocket universes for
which the time of formation
\beq
  t_i \le t_c \ , 
  \label{eq:8.2}
\eeq
then the mean value of $t_{\rm rel}$ is equal to $T/2$ for any
$t_c$.  The truncation method of Eq.~(\ref{eq:8.1}) leads to the
youngness paradox, in which the probability sample is strongly
dominated by universes that are extremely young, while the
truncation method of Eq.~(\ref{eq:8.2}) does not.

At this point, I have to admit that I do not understand how to
resolve the ambiguities associated with this toy model.  It is
conceivable that there is no meaningful method of regularization,
and that $t_{\rm rel}$ is somehow not susceptible to
probabilistic predictions.  It is also conceivable that there is
something wrong with either the truncation (\ref{eq:8.1}) or
(\ref{eq:8.2}) or both, and that a correct analysis would lead to
a unique probability calculation.  It is also conceivable that
the regularization has to be specified as part of the theory, so
that the truncations (\ref{eq:8.1}) and (\ref{eq:8.2}) represent
two distinct theories, each of which is logically consistent.

\section{An Alternative Probability Prescription}
\setcounter{equation}{0}

Since the probability measure depends on the method used to
regulate the infinite spacetime of eternal inflation, we are not
forced to accept the consequences of the synchronous gauge
probabilities.  A very attractive alternative has been proposed
by Vilenkin \cite{vilenkin-proposal}, and developed further by
Vanchurin, Vilenkin, and Winitzki \cite{vvw}.  This procedure is,
roughly speaking, analogous to the truncation of
Eq.~(\ref{eq:8.2}). 

The key idea of the Vilenkin proposal is to define probabilities
within a single pocket universe (which he describes more
precisely as a connected, thermalized domain).  Thus, unlike the
synchronous gauge method, there is no comparison between old
pocket universes and young ones.  To justify this approach it is
crucial to recognize that each pocket universe is infinite, even
if one starts the model with a finite region of de Sitter space. 
The infinite volume arises in the same way as it does for the
special case of Coleman-de Luccia bubbles
\cite{coleman-deluccia}, the interior of which are open
Robertson-Walker universes.  From the outside one often describes
such bubbles in a coordinate system in which they are finite at
any fixed time, but in which they grow without bound.  On the
inside, however, the natural coordinate system is the one that
reflects the intrinsic homogeneity, in which the space is
infinite at any given time.  The infinity of time, as seen from
the outside, becomes an infinity of spatial extent as seen on the
inside.  Thus, at least for continuously variable parameters, a
single pocket universe provides an infinite sample space which
can be used to define probabilities.  The second key idea of
Vilenkin's method is to use the inflaton field itself as the time
variable, rather than the synchronous time variable discussed in
the previous section.

This approach can be used, for example, to discuss the
probability distribution for $\Omega$ in open inflationary
models, or to discuss the probability distribution for some
arbitrary field that has a flat potential energy function.  If,
however, the vacuum has a discrete parameter which is homogeneous
within each pocket universe, but which takes on different values
in different pocket universes, then this method does not apply. 

\begin{figure}[ht]
\centerline{\epsfbox{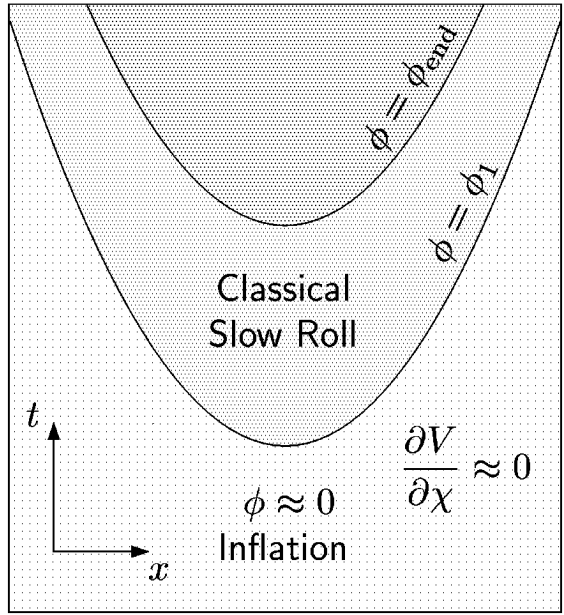}}
\caption{A schematic picture of a pocket universe, illustrating
Vilenkin's proposal for the calculation of probabilities.} 
\label{vilenkin-space}
\end{figure}

The proposal can be described in terms of
Fig.~\ref{vilenkin-space}.  We suppose that the theory includes
an inflaton field $\phi$ of the new inflation type, and some set
of fields $\chi_i$ which have flat potentials.  The goal is to
find the probability distribution for the fields $\chi_i$.  We
assume that the evolution of the inflaton $\phi$ can be divided
into three regimes, as shown on the figure.  $\phi < \phi_1$
describes the eternally inflating regime, in which the evolution
is governed by quantum diffusion.  For $\phi_1 < \phi < \phi_{\rm
end}$, the evolution is described classically in a slow-roll
approximation, so that $\dot \phi \equiv {\rm d} \phi / {\rm d}
t$ can be expressed as a function of $\phi$.  For $\phi >
\phi_{\rm end}$ inflation is over, and the $\phi$ field no longer
plays an important role in the evolution.  The $\chi_i$ fields
are assumed to have a finite range of values, such as angular
variables, so that a flat probability distribution is
normalizable.  They are assumed to have a flat potential energy
function for $\phi > \phi_{\rm end}$, so that they could settle
at any value.  They are also assumed to have a flat potential
energy function for $\phi < \phi_1$, although they might interact
with $\phi$ during the slow-roll regime, however, so that they
can affect the rate of inflation. 

Since the potential for the $\chi_i$ is flat for $\phi < \phi_1$,
we can assume that they begin with a flat probability
distribution $P_0(\chi_i) \equiv P(\chi_i, \phi_1)$ on the $\phi
= \phi_1$ hypersurface.  If the kinetic energy function for the
$\chi_i$ is of the standard form, we take $P_0(\chi_i) = const$. 
If, however, the kinetic energy is nonstandard,
\beq
  {\cal L}_{\rm kinetic} = g^{ij}(\chi) \partial_\mu \chi_i
     \partial^\mu \chi_j \ ,
  \label{eq:9.1}
\eeq
as is plausible for a field described in angular variables, then
the initial probability distribution is assumed to take the
reparameterization-invariant form
\beq
  P_0(\chi_i) \propto \sqrt{ \det g}  \ .
  \label{eq:9.2}
\eeq
During the slow-roll era, it is assumed that the $\chi_i$ fields
evolve classically, so one can calculate the number of e-folds of
inflation $N(\chi_i)$ as a function of the final value of the
$\chi_i$ (i.e., the value of $\chi_i$ on the $\phi = \phi_{\rm
end}$ hypersurface).  One can also calculate the final values
$\chi_i$ in terms of the initial values $\chi_i^0$ (i.e., the
value of $\chi_i$ on the $\phi=\phi_1$ hypersurface). One then
assumes that the probability density is enhanced by the volume
inflation factor $e^{3 N (\chi_i)}$.  The evolution from
$\chi_i^0$ to $\chi_i$ results in a Jacobian factor.  The
(unnormalized) final probability distribution is thus given by
\beq
  P(\chi_i, \phi_{\rm end}) = P_0(\chi_i^0) e^{3 N (\chi_i)} \,
     \det {\partial \chi_j^0 \over \partial \chi_k} \ .
  \label{eq:9.3}
\eeq
Alternatively, if the evolution of the $\chi_i$ during the
slow-roll era is subject to quantum fluctuations, Ref.~\cite{vvw}
shows how to write a Fokker-Planck equation which is equivalent
to averaging the result of Eq.~(\ref{eq:9.3}) over a collection of
paths that result from interactions with a noise term.

The Vilenkin proposal sidesteps the youngness paradox by defining
probabilities by the comparison of volumes within one pocket
universe.  The youngness paradox, in contrast, arose when one
considered a probability ensemble of all pocket universes at a
fixed value of the synchronous gauge time coordinate---an
ensemble that is overwhelmingly dominated by very young pocket
universes.

The proposal has the drawback, however, that it cannot be used to
compare the probabilities of discretely different alternatives. 
Furthermore, although the results of this method seem
reasonable, I do not at this point find them compelling.  That
is, it is not clear what principles of physics or probability
theory ensure that this particular method of regularizing the
spacetime is the one that leads to correct predictions.  Perhaps
there is no way to answer this question, so we may be forced to
accept this proposal, or something similar to it, as a postulate.

\section{Probabilities with only one universe?}

In discussing a probabilistic approach to cosmology, we need to
know whether it makes sense to talk about a probability
distribution for a cosmic parameter such as $\Omega$, for which
we have only one example to measure.  I have certainly heard more
than one physicist say that he or she doesn't think that one can
meaningfully talk about probabilities for an experiment that can
be done only once.  The notion that probability requires
repetition is very widespread, and I am sure that it is
incorporated into many books about probability theory. 
Nonetheless, I would like to argue that repetition is not at all
necessary to make use of probability theory.  Instead, I will
argue that probability is meaningful whenever one has a {\it
strong} probabilistic prediction, by which I mean a prediction
that the probability for some discernible event is either very
close to zero or very close to one.

Thus, if a cosmological theory predicts a probability
distribution for $\Omega$ which is reasonably flat, then there is
no strong prediction, and the implications of the theory for
$\Omega$ do not provide a way of testing the theory.  However, if
the theory predicts that the probability of $\Omega$ lying
outside the range of 0.99 to 1.01 is $10^{-6}$, then I would
claim that the prediction is meaningful and can be used to test
the theory.

My point of view can be explained most easily by considering coin
flips.  If a flip of an unbiased coin is repeated 20 times, the
probability of getting 20 successive heads is a very small
number, about $10^{-6}$.  This is an example of what I call a
strong prediction.  Many common examples of strong predictions
involve repetition.  However, if 20 unbiased coins are flipped
simultaneously in a single experiment, the probability that all
will come up heads is identical, about $10^{-6}$.  Since the
probability of these two results---20 successive heads or 20
simultaneous heads---are both equally small, I would draw the
obvious conclusion that we should be equally surprised if either
result occurred.  It does not matter that the first result
involved repetition, while the second did not.  Some might argue
that the 20 simultaneous coin flips involved the replication of
identical experiments, even if they were not performed in
succession, so I will take the analogy one step further.  Suppose
we constructed a roulette wheel that was so finely ruled that
the probability of the ball landing on 0 was only $10^{-6}$. 
Again, we should be just as surprised if this result occurred as
we would be if 20 successive coins landed heads.

Similarly, if our cosmological theory predicted that the
probability of $\Omega$ lying outside the range of 0.99 to 1.01
is $10^{-6}$, we should be just as surprised if this outcome
occurred as we would be if 20 consecutive coins came up heads. 
In both cases, we would have good cause to question the
assumptions that went into calculating the prediction.

\section{Conclusion}
\setcounter{equation}{0}

In this paper I have summarized the workings of inflation, and
the arguments that strongly suggest that our universe is the
product of inflation.  I argued that inflation can explain the 
size, the Hubble expansion, the homogeneity, the isotropy, and the
flatness of our universe, as well as the  absence of magnetic
monopoles, and even the characteristics of the nonuniformities. 
The detailed observations of the cosmic background radiation
anisotropies continue to fall in line with inflationary
expectations, and the evidence for an accelerating universe fits
well with the inflationary preference for a flat universe.

Next I turned to the question of eternal inflation, claiming
that essentially all inflationary models are eternal. In my opinion
this makes inflation very robust: if it starts anywhere,
at any time in all of eternity, it produces an infinite number of
pocket universes.  Eternal inflation has the very attractive
feature, from my point of view, that it offers the possibility of
allowing unique predictions even if the underlying string theory
does not have a unique vacuum.  I have also emphasized, however,
that there are important problems in understanding the
implications of eternal inflation.  First, there is the problem
that we do not know how to treat the situation in which the
scalar field climbs upward to the Planck energy scale.  Second,
the definition of probabilities in an eternally inflating
spacetime is not yet a closed issue, although important progress
has been made.  And third, I might add that the entire present
approach is at best semiclassical.  A better treatment may not be
possible until we have a much better handle on quantum gravity,
but eventually this issue will have to be faced.

\section*{Acknowledgments}

The author particularly thanks Andrei Linde, Alexander Vilenkin,
Neil Turok, and other participants in the Isaac Newton Institute
programme {\it Structure Formation in the Universe} for very
helpful conversations.  This work is supported in part by funds
provided by the U.S. Department of Energy (D.O.E.) under
cooperative research agreement \#DF-FC02-94ER40818, and in part
by funds provided by NM Rothschild \& Sons Ltd and by the EPSRC.

\newcommand{\jf}{\it}  
\newcommand{\jt}{\/}   
\newcommand{\VPY}[3]{{\bf #1}, #2 (#3)}  
\newcommand{\ispace}{\thinspace}   

\let\U=\.
\def\.{.\nobreak\ispace\ignorespaces}

\newcommand{\IJMODPHYS}[3]{{\jf Int. J. Mod. Phys.\jt} \VPY{#1}{#2}{#3}}
\newcommand{\JETP}[3]{{\jf JETP Lett.\jt} \VPY{#1}{#2}{#3}}
\newcommand{\MPL}[3]{{\jf Mod. Phys. Lett.\jt} \VPY{#1}{#2}{#3}}
\newcommand{\NC}[3]{{\jf Nuovo Cim.\jt} \VPY{#1}{#2}{#3}}
\newcommand{\NP}[3]{{\jf Nucl. Phys.\jt} \VPY{#1}{#2}{#3}}
\newcommand{\PHYREP}[3]{{\jf Phys. Rept.\jt} \VPY{#1}{#2}{#3}}
\newcommand{\PL}[3]{{\jf Phys. Lett.\jt} \VPY{#1}{#2}{#3}}
\newcommand{\PRD}[3]{{\jf Phys. Rev. D\jt} \VPY{#1}{#2}{#3}}
\newcommand{\PRL}[3]{{\jf Phys. Rev. Lett.\jt} \VPY{#1}{#2}{#3}}
\newcommand{\PTRSLA}[3]{{\jf Phil. Trans. R. Soc. Lond.\jt\ A}
\VPY{#1}{#2}{#3}}
\newcommand{\ZhETF}[3]{{\jf Zh. Eksp. Teor. Fiz.\jt} \VPY{#1}{#2}{#3}}

\end{document}